\documentclass[aps,preprint,showpacs,preprintnumbers,amsmath,amssymb,floatfix]{revtex4-1}
\usepackage{graphicx}
\usepackage{amsfonts}
\usepackage{CJK}

\begin{document}

\title{Interfacial thermal conduction and negative temperature jump in one-dimensional lattices}

\author{Xiaodong Cao}
\author{Dahai He}\email{dhe@xmu.edu.cn}

\affiliation{Department of Physics and Institute of Theoretical Physics and Astrophysics, Xiamen University, Xiamen 361005, Fujian, China }

\date{\today}

\begin{abstract}
We study the thermal boundary conduction in one-dimensional harmonic and $\phi^{4}$ lattices, both of which consist of two segments coupled by a harmonic interaction. For the ballistic interfacial heat transport through the harmonic lattice,
we use both theoretical calculation and molecular dynamics simulation to study the heat flux and
temperature jump at the interface as to gain insights of the Kapitza resistance at the atomic scale. In the weak coupling regime, the heat current is proportional to the square of the
coupling strength for the harmonic model as well as anharmonic models. Interestingly, there exists a negative temperature jump between the interfacial particles in particular parameter regimes. A nonlinear response of the boundary temperature jump to the externally applied temperature difference in the $\phi^{4}$ lattice is observed. To understand the anomalous result, we then extend our studies to a model in which the interface is represented by a relatively small segment with gradually changing spring constants, and find that the negative temperature jump still exist. Finally, we show that the local velocity distribution at the interface is so close to the Gaussian distribution that the existence/absence of local equilibrium state seems unable to determine by numerics in this way.
\end{abstract}


\maketitle
\section{Introduction}\label{INTRODUCTION}

Since the pioneering observation between liquid helium and a metal~\cite{KAPITZA1}, thermal boundary resistance, namely, Kapitza resistance, has been extensively studied theoretically and experimentally~\cite{KAPITZA2}. With the rapid development of modern electronic technology, there has been much need and interest in understanding the fundamental nature of thermal boundary conductance since it has been a significant obstacle in designing the micro- or nano- scale electronic chips. Two phenomenological models, the acoustic mismatch model \cite{ACCUSTIC} and diffuse mismatch model \cite{KAPITZA2}, have been proposed to study the mechanism of the thermal boundary conductance. However, due to the neglecting of atomic details at the interface, they both offer limited accuracy, particularly, for nanoscale interfacial resistance~\cite{ACCUSTIC2}. To understand the mechanism of thermal boundary conductance at the atomic level,
 many studies have been done in one-dimensional lattices via different methods~\cite{ACCUSTIC2,SBM1,HE1}. Most of the previous studies focus on the effect of the interface on the steady-state heat flux and little attention
has been paid to the temperature jump between the interface from the atomic viewpoint.

On the other hand, heat conduction in low-dimensional dynamical systems has become the subject of a large number of theoretical and
experimental studies in recent years
\cite{HEAT-REVIEW1,HEAT-REVIEW2,HEAT-NANO1}.
An exact approach to  interacting Hamiltonian systems is so far unavailable except for harmonic crystals.
A meaningful  definition of local temperature depends on the local thermal equilibrium and it is difficult
to give a microscopic derivation of
the condition in general~\cite{HEAT-REVIEW3}. With the usual definition of local temperature i.e.,
the mean local kinetic energy, the temperature profile
may show some unexpected features, such as the temperature oscillations in the steady state of alternate
mass hard particle gas \cite{TEM-GAS1}, in the
Fermi-Pasta-Ulam chain \cite{TEM-FPU} and in harmonic chain with alternating mass \cite{TEM-HAR}.

In the present study, we study the heat flux and temperature jump at the interface as to gain insights of the Kapitza resistance at the atomic scale via theoretical calculations and molecular dynamics simulations. We find that there exists a negative temperature jump between the interfacial particles in particular parameter regimes. A nonlinear response of the boundary temperature jump to the externally applied temperature difference in the $\phi^{4}$ lattice is observed.
Note that, although the \emph{interface} between two segments is not well defined in one-dimensional Hamiltonian systems,
our studies can give some insights into the thermal boundary resistance in real systems.

The paper is organized as follows. In Sec.~\ref{sec2} we define the model and give the methods for theoretical
calculations and molecular dynamics simulations.
In Sec.~\ref{sec3} we demonstrate the existence of negative temperature jump in both the harmonic and $\phi^4$ model. Finally, we give a brief summary and discussion in Sec.~\ref{sec4}.
\section{Model and Methods}\label{sec2}
We study the non-equilibrium steady state of a one dimensional chain consisting of two coupled lattice,
\begin{eqnarray}\label{HFUL}
H=H_{L}+H_{R}+\frac{1}{2}k_{c}\left(x_{N/2}-x_{N/2+1}\right)^{2}.
\end{eqnarray}
The Hamiltonian for the left and right segments are given by
\begin{eqnarray}
&&H_{L}=\sum_{i=1}^{N/2}\left(\frac{p_{i}^{2}}{2m}+\frac{f_{L}}{2}{x_{i}^{2}}+\frac{\lambda_{L}}{4}x_{i}^{4}\right)+
\frac{k_{L}}{2}\sum_{i=0}^{N/2-1}\left(x_{i}-x_{i+1} \right)^{2}, \\ \nonumber
&&H_{R}=\sum_{i=N/2+1}^{N}\left(\frac{p_{i}^{2}}{2m}+\frac{f_{R}}{2}x_{i}^{2}+\frac{\lambda_{R}}{4}x_{i}^{4}\right)+
\frac{k_{R}}{2}\sum_{i=N/2+1}^{N}\left(x_{i}-x_{i+1}\right)^{2}, \nonumber
\end{eqnarray}
where $x_{i}$ denotes the displacement of the $i$-th particle from its equilibrium position. Fixed boundary conditions are taken, i.e.,
$x_{0}=x_{N+1}=0$. The particle $1$ and $N$ at the two ends are connected to the heat baths at
temperature $T_{L}$ and $T_{R}$, respectively. The heat baths are modeled by the Langevin equations corresponding
to Ohmic baths, i.e., the self energy of the baths are $\Sigma(\omega)=i\gamma\omega$ \cite{HEAT-REVIEW2}.

When $\lambda_{L}=\lambda_{R}=0$, the on-site potential and inter-particle interaction are all quadratic.
In the classical limitation, the steady heat current from
left to right reservoir can be given by the Langevin equations and Green's function (LEGF) method \cite{HEAT-REVIEW2,DHAR2}
\begin{eqnarray}\label{current}
J=\frac{k_{B}(T_{L}-T_{R})}{\pi}\int_{-\infty}^{\infty}d\omega \mathrm{Tr}
\left[G^{+}_{S}(\omega)\Gamma_{L}(\omega)G^{-}_{S} \Gamma_{R}(\omega)\right]
\end{eqnarray}
with
\begin{eqnarray}
&& G^{\pm}_{S}(\omega)=\frac{1}{\left[-\omega^{2}M_{S}+K_{S}
-\Sigma_{L}^{\pm}(\omega)-\Sigma^{\pm}_{R}(\omega)\right]}, \\
&& \Gamma_{L,R}(\omega)=\mathrm{Im}\left(\Sigma^{+}_{L,R}(\omega)\right),
\end{eqnarray}
where $M_{S}$ and $K_{S}$ denote the mass matrix and force constant matrix of the system.
Note that $G_{S}^{\pm}, \Sigma_{L,R}^{\pm}$ are all $N\times N$ matrices for one-dimensional systems.
The only non-zero element of $\Sigma^{\pm}_{L,R}$ are respectively $[\Sigma_{L}^{\pm}]_{1,1}=\Sigma=i\gamma\omega$
and $[\Sigma_{L}^{\pm}]_{N,N}=\Sigma=i\gamma\omega$. $\gamma$ is the coupling strength of the first
and $N$-th particle to the left and right reservoirs, respectively.
 The velocity-velocity correlation and position-velocity correlation are:
\begin{eqnarray}\label{tem}
&&K=\langle\dot{\tilde{X}}_{S}\dot{\tilde{X}}_{S}^{T}\rangle=
\frac{k_{B}T_{L}}{\pi}\int_{-\infty}^{\infty}d\omega\omega G^{+}_{S}(\omega)\Gamma_{L}(\omega)G^{-}_{S}
(\omega) \\ \nonumber && +\frac{k_{B}T_{R}}{\pi}\int_{-\infty}^{\infty}d\omega\omega G^{+}_{S}(\omega)
\Gamma_{R}(\omega)G^{-}_{S}(\omega),
\end{eqnarray}
\begin{eqnarray}\label{local_current}
&&C=\langle\tilde{X}_{S}\dot{\tilde{X}}_{S}^{T}\rangle= \frac{ik_{B}T_{L}}{\pi}\int_{-\infty}^{\infty}
d\omega G^{+}_{S}(\omega)\Gamma_{L}(\omega)G^{-}_{S}(\omega)
 \\ \nonumber && +\frac{ik_{B}T_{R}}{\pi}\int_{-\infty}^{\infty}d\omega G^{+}_{S}(\omega)\Gamma_{R}
 (\omega)G^{-}_{S}(\omega).
\end{eqnarray}
The correlation function $K$ can be used to define local energy density which can in turn be used to define
the local temperature,
i.e.,
\begin{equation}\label{temperature}
T_{i}=mK_{i,i},
\end{equation}
and $C$ gives the local heat current density \cite{KC1,KC2,KC3,KC4}. We integrate Eq.~\eqref{current} and Eq.~\eqref{tem} numerically to obtain the
steady state heat current and local temperature, for which the rectangular method is used~\cite{Numerical}. We also verify that the local current, obtained by integrating Eq.~\eqref{local_current} numerically, is the same along the chain, which is one of the properties in the steady state.

When $\lambda_{L},\lambda_{R}\neq0$, we apply the non-equilibrium molecular dynamics simulation(NEMD) to the system, for which the Langevin heat baths are used at the two ends of the chain. The equations of motion are given by
\begin{eqnarray}\label{eom}
m\ddot{x}_{i}=-\frac{\partial H}{\partial x_{i}}-\gamma_{i}\dot{x}_{i}+\eta_{i},
\end{eqnarray}
where $\gamma_{i}=\gamma(\delta_{1,i}+\delta_{N,i})$ and $\eta_{i}=\eta_{L}\delta_{1,i}+\eta_{R}\delta_{N,i}$.
The noise terms $\eta_{L,R}$ denotes a Gaussian white
noise with zero mean and variance of $2\gamma k_{B}T_{L,R}$. The local heat flux is given by
$j=\langle F(x_{i+1}-x_{i})v_{i+1}\rangle$, where $F(x)=-V'(x)$ and
the notion $\langle ...\rangle$ denotes a steady-state average. The equations of motion (Eq.~\eqref{eom}) are integrated by using a second-order Stochastic Runge-Kutta algorithm~\cite{SRK2}.  At steady states, the numerically
computed local heat flux is always constant along the chain, and the
local temperature is defined as $T_{i}=m\langle \dot x_{i}^{2}\rangle$. To compute the boundary temperature jump,
 i.e., $\Delta T_{b}=T_{N/2}-T_{N/2+1}$,
the relaxation and average time must  be both long enough. In what follows we set $m=1, k_{B}=1$, and $\gamma=1$.
\section{Results and Discussions}\label{sec3}
\begin{figure}[h!]
\includegraphics[width=1\linewidth]{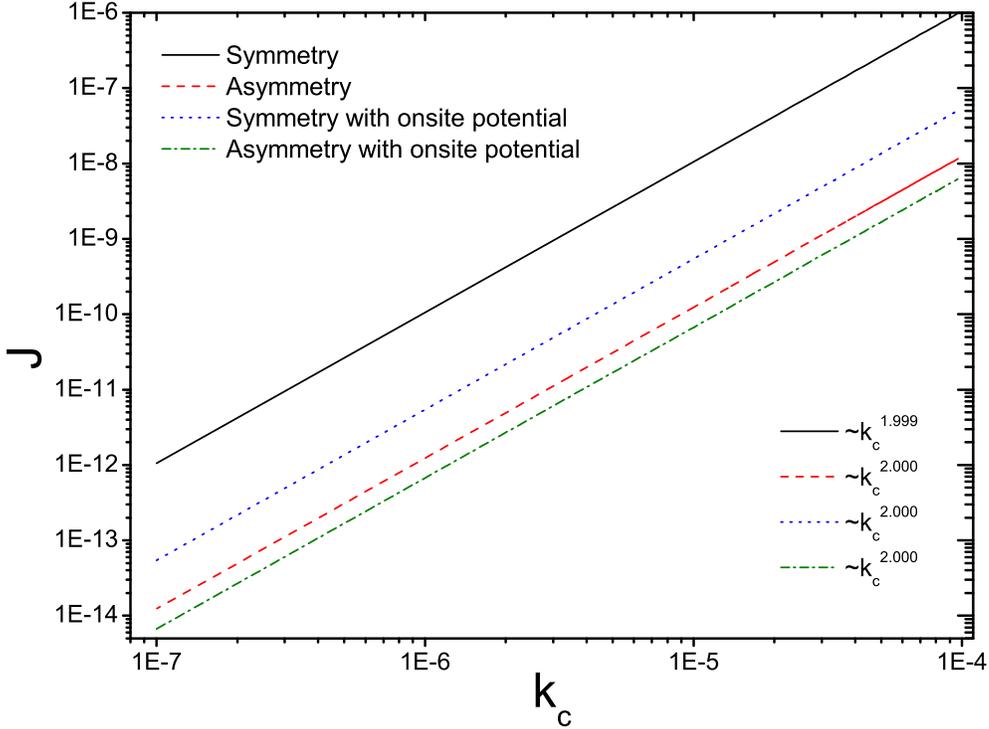}
\caption{Heat flux as a function of interfacial coupling $k_{c}$ via the LEGF approach as $k_{c}$ approaches to zero. Symmetry: $k_{L}=k_{R}=1,f_{L}=f_{R}=0$; Asymmetry: $k_{L}=1,k_{R}=2,f_{L}=f_{R}=0$;
Symmetry with on-site potential: $k_{L}=1,k_{R}=1,f_{L}=f_{R}=2$. Asymmetry with on-site potential: $k_{L}=1,k_{R}=2,f_{L}=f_{R}=2$. For all cases, we set $T_{L}=2, T_{R}=1$ and $N=64$. } \label{Fig1}
\end{figure}
For many devices of several segments, interfacial coupling is pretty weak, which indicates that $k_{c}$ is far less than $k_{L}$ and $k_{R}$ in our model. So it is desirable to study the thermal transport through atomic chains in the weak coupling regime. It has been shown that~\cite{SCPT1} the heat current is proportional to the square of the coupling strength in one-dimensional weakly-coupled chain with the Morse on-site potential by a phenomenological analysis. Does this square-law relation between heat current and coupling strength is still valid in the weak coupling limit when the anharmonic on-site potential is absent? As shown in Fig.\ref{Fig1}, we plot the heat current as a function of the coupling strength in the weak
coupling limit by integrating Eq.~(\ref{current}) numerically. It turns out that the square law relation is still valid when the anharmonicity is absent. Further
more, the square law relation still holds when the system consists of symmetrical/asymmetrical segments with/without an on-site potential.
\begin{figure}[h!]
\includegraphics[width=1\linewidth]{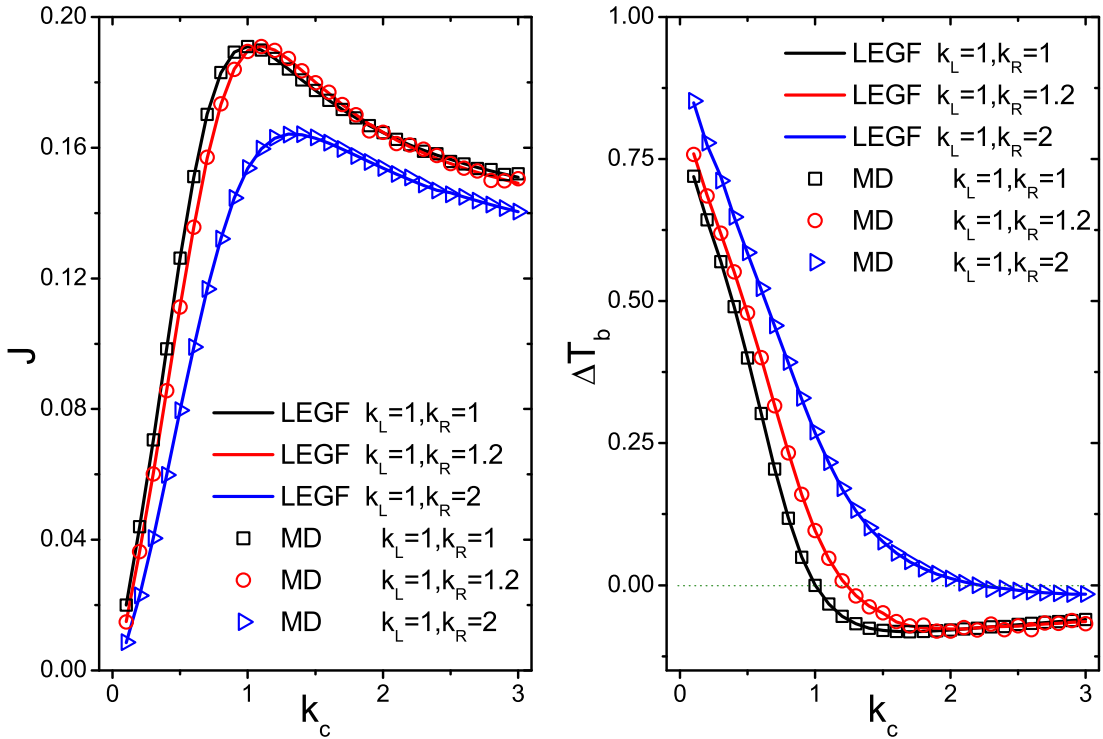}
\caption{The heat flux (left) and temperature jump between the
$N/2$-th and $(N/2+1)$-th particles (right) as a function of $k_{c}$ via both the LEGF approach and MD simulations.
Here $f_{L}=f_{R}=0,T_{L}=2,T_{R}=1,\lambda_{L}=\lambda_{R}=0$, and $N=64$. } \label{Fig2}
\end{figure}
Fig.~\ref{Fig2} shows the steady-state heat current and the boundary temperature jump as a function of the coupling strength $k_{c}$.
The reason to carry out both theoretical calculation and NEMD simulation is to verify that the results we obtained are from physical
reasons rather than uncertain numerical reasons because the temperature jump between the $N/2$-th particle and the $(N/2+1)$-th
particle requires a highly accurately performed simulation for its sensitive to heat fluctuation when $k_{c}$ approaches to
$k_{R}$. By inspecting the figure, we can see that theoretical calculations and MD simulations agree well with each other. The heat current increases at first, then arrives at a maximum value, and then slightly
decreases with the increase of $k_{c}$. As depicted in Fig.~\ref{Fig2}, the maximum heat current occurs at $k_{c}=2k_{L}k_{R}/(k_{L} +k_{R})$, which agrees with the result obtained in~\cite{SBM1} by the scattering boundary method. Furthermore, both theoretical calculations and MD simulations indicate that there is a negative boundary temperature jump, i.e., $\Delta T_b<0$, when $k_{c}$ approaches to $k_{R}$. The word \textquotedblleft negative" is in contrast with \textquotedblleft normal\textquotedblright heat conduction
that the direction of the heat flow is from hot to cold regions. In fact, similar negative temperature jumps occurs in several
systems, for example, the temperature jump between the second and third particle, and between $(N-2)$-th and $(N-1)$-th particle
in the uniform harmonic chain \cite{HEAT-REVIEW1} coupled with reservoirs, and the temperature oscillations in the steady state of hard
particle gas  \cite{TEM-GAS1}, the Fermi-Pasta-Ulam chain \cite{TEM-FPU}, the harmonic chain \cite{TEM-HAR} with alternating mass.
\begin{figure}[h!]
\centering
\includegraphics[width=1\linewidth]{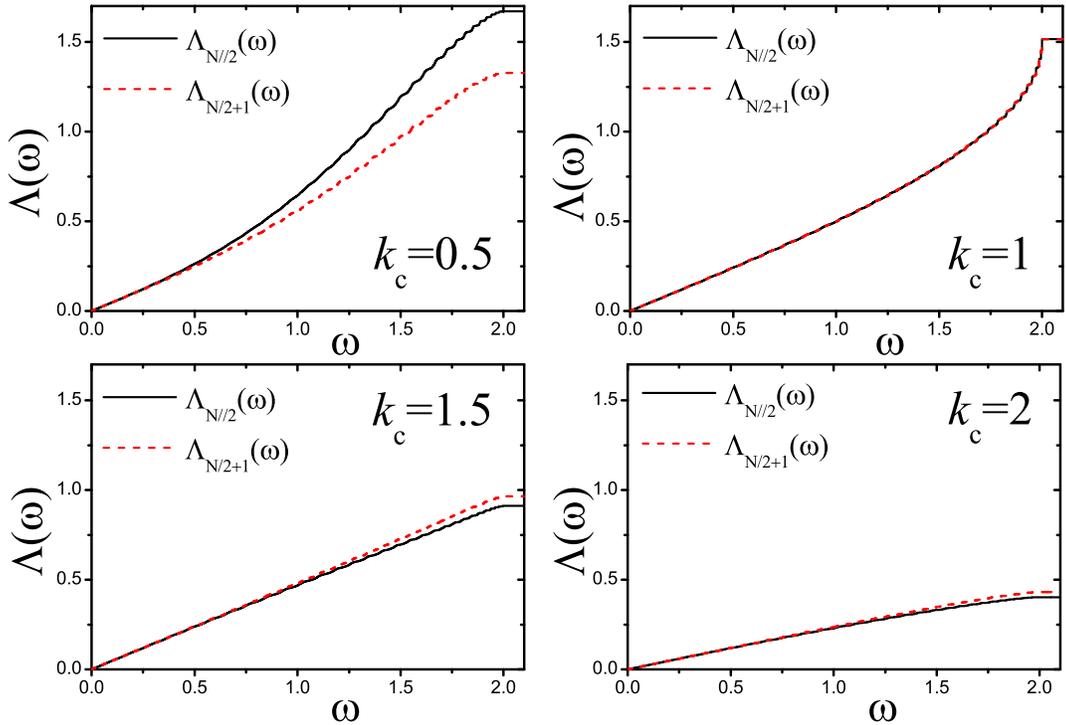}
\caption{The contribution of normal modes to local temperature at the interface. Here $k_{L}=k_{R}=1$, $f_{L}=f_{R}=0$, $\lambda_{L}=\lambda_{R}=0$, $T_{L}=2, T_{R}=1$ and $N=64$.} \label{Fig3}
\end{figure}
To understand the negative temperature jump at the interface, we need to inspect the concept of the local temperature further. The local temperature of the $i$-th particle can be written as $T_{i}=\Lambda_{i}(\omega_{max})$,
with
\begin{eqnarray}\label {Lambda}
\Lambda_{i}(\omega)=2\int_{0}^{\omega}d\omega'\left(
\frac{k_{B}T_{L}}{\pi}\omega' G^{+}_{S}(\omega')\Gamma_{L}(\omega')G^{-}_{S}
(\omega')+\frac{k_{B}T_{R}}{\pi}\omega' G^{+}_{S}(\omega')
\Gamma_{R}(\omega')G^{-}_{S}(\omega')\right),
\end{eqnarray}
and $\omega_{max}$ is the top boundary of the phonon spectra. The kinetic energy of a particle
gets contributions from all the modes, and the net result depends on the distribution of
energy in the different modes. As shown in Fig.~\ref{Fig3}, we plot the contribution of normal modes to the local temperature
for the $(N/2)$-th and $(N/2+1)$-th particles by integrating Eq.~\eqref{Lambda} numerically. As we can see, \textquotedblleft equipartition" among phonon modes, i.e., each normal mode shares the same average kinetic energy, is not satisfied for $k_{c}=0.5$ and $k_{c}=1$
shown by the nonlinear behaviors of $\Lambda_{i}(\omega)$($i=N/2,N/2+1$) in the high frequencies region. Surprisingly,
for the case of $k_{c}=1.5$, $\Lambda_{i}(\omega)$ exhibit almost linear behavior with the increasing of $\omega$, indicating that the contribution to the local temperature from possible phonon modes are closely equivalent.
Comparing with $k_{c}=0.5$, we can see that the high frequency normal modes are suppressed more dramatically than the low frequency normal modes for $k_{c}=1.5$ and $k_{c}=2$, and the turning of $\Lambda_{N/2}$ and $\Lambda_{N/2+1}$ in the high frequencies region indicates the negative temperature jump between the $N/2$-th and $(N/2+1)$-th particle.

\begin{figure}[h!]
\centering
\includegraphics[width=1\linewidth]{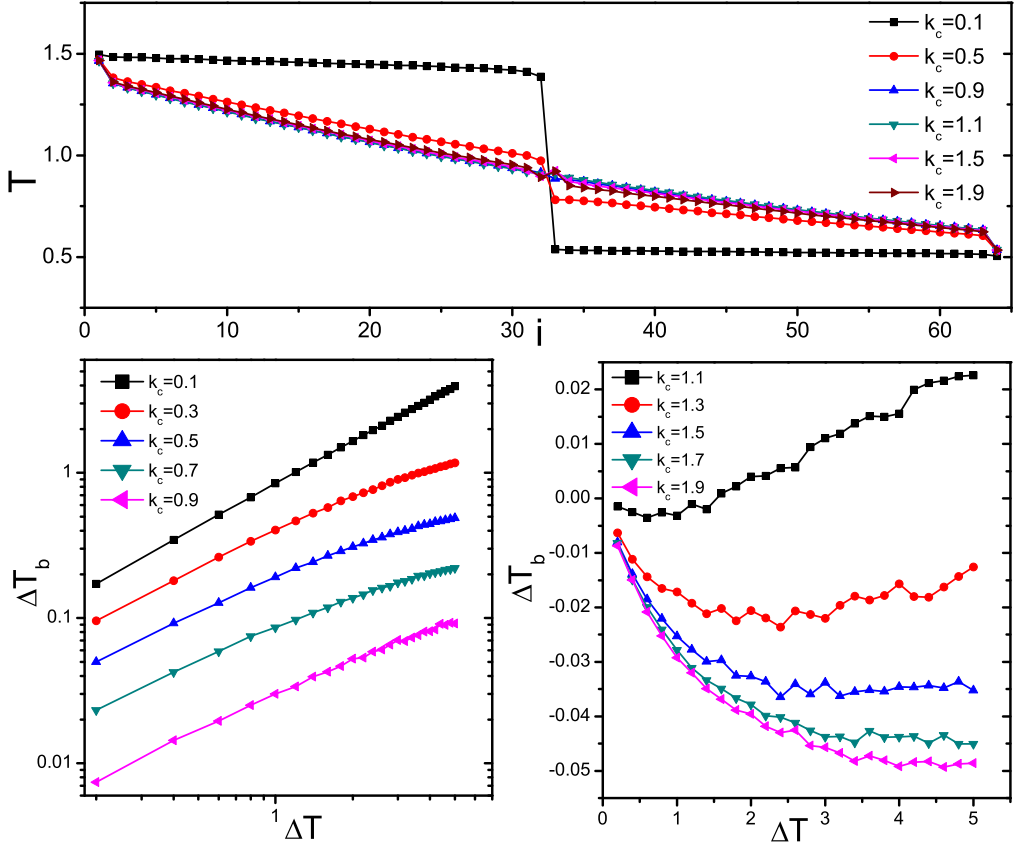}
\caption{$\Delta T_{b}$ as a function of $\Delta T$ for different $k_{c}$ in the $\phi^{4}$ lattice , with
$T_{L}=T_{R}+\Delta T, T_{R}=0.5, \lambda_{L}=\lambda_{R}=1, k_{L}=k_{R}=1$ and  $N=64$.} \label{Fig4}
\end{figure}
It would be interesting to see if the negative temperature jump is an artificial effect due to the integrability of the harmonic system. Thus we conduct similar studies in the $\phi^{4}$ lattice, which has additional nonlinear on-site potential on each site in comparison with the harmonic system. We plot the boundary temperature jump
$\Delta T_{b}$ as a function of the external temperature difference $\Delta T=T_{L}-T_{R}$ and some typical
temperature profiles in Fig. \ref{Fig4}. One can see that the boundary temperature jump is
proportional to $\Delta T$ for $k_{c}<1$ and proportional to $(-\Delta T)$ for $k_{c}>1$ when $\Delta T$ is small, which are typical linear-response behaviors shown in harmonic models. With the increasing of $\Delta T$, the linear behavior of $\Delta T_{b}$ no longer holds for the $\phi^{4}$ lattice. Note that negative temperature jump occurs when $k_{c} \geq 1.3$ and the absolute value of $\Delta T_{b}$ nonlinearly increases as $\Delta T$ increases.

\begin{figure}[h!]
\centering
\includegraphics[width=1\linewidth]{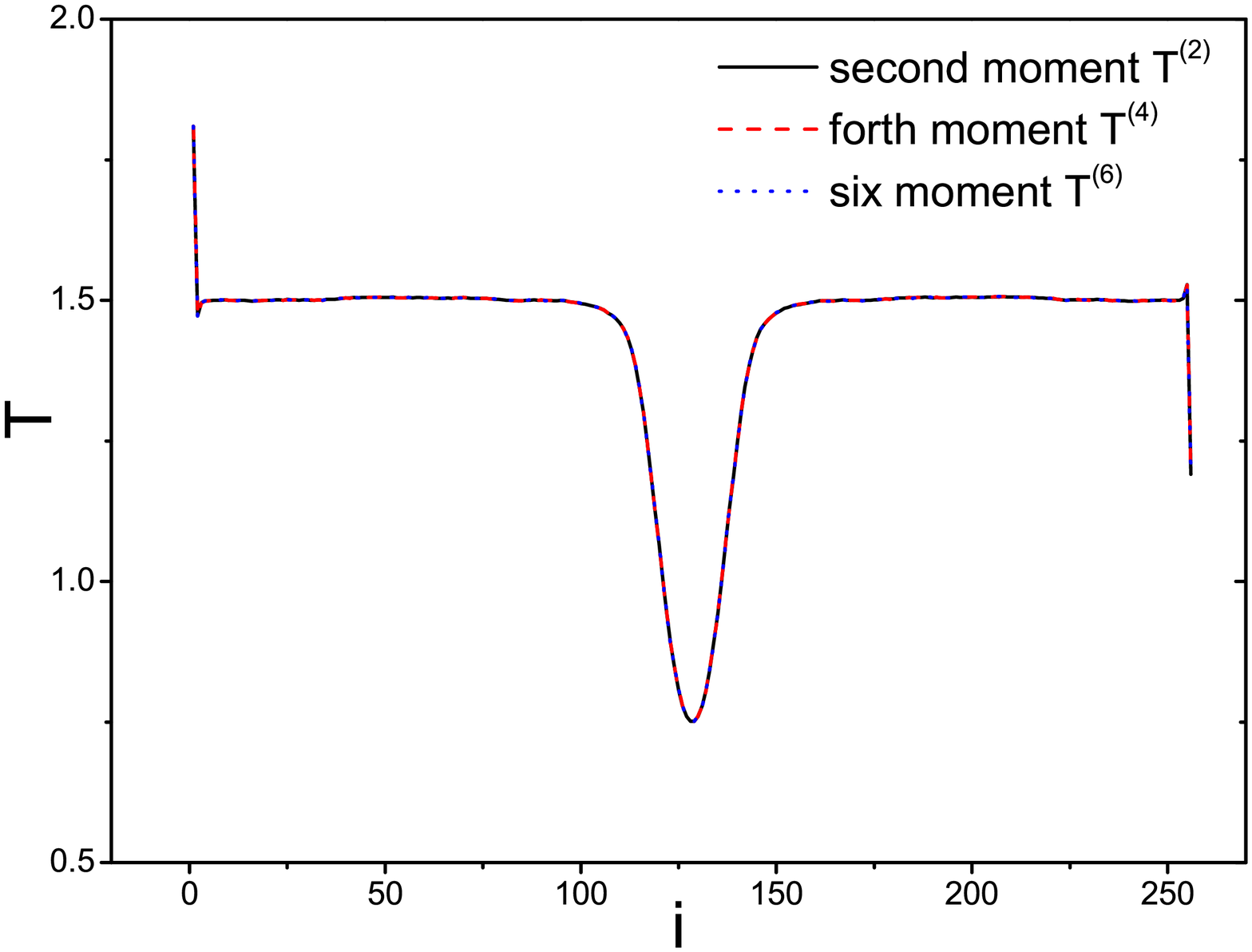}
\caption{The temperature profile of the harmonic chain with an interfacial junction, whose spring constants are smoothly varied.  The distribution of spring constants for the whole system is given as follows: $k_i=k_{L}=1$ for $1\leq i\leq 7N/16$; $k_i=\exp{(-(i-N/2)^{2}/50)}+1$ for $7N/16 < i \leq 9N/16$; and $k_i=k_{R}=1$ for $9N/16< i\leq (N-1)$, where $i$ is the index of particle number and $N=256$. The first three even moments of velocity are given by $T_{i}^{(2)}=m\langle \dot{x}_{i}^{2}\rangle$ for the second moment, $T_{i}^{(4)}=m\left(\langle\dot{x_{i}}^{4}\rangle/3\right)^{1/2}$ for the fourth moment and $T_{i}^{(6)}=m\left(\langle\dot{x_{i}}^{6}\rangle/15\right)^{1/3}$ for the sixth
moment, respectively. Here $T_{L}=2, T_{R}=1$, and $N=256$.} \label{Fig5}
\end{figure}
\begin{figure}[h!]
\centering
\includegraphics[width=1\linewidth]{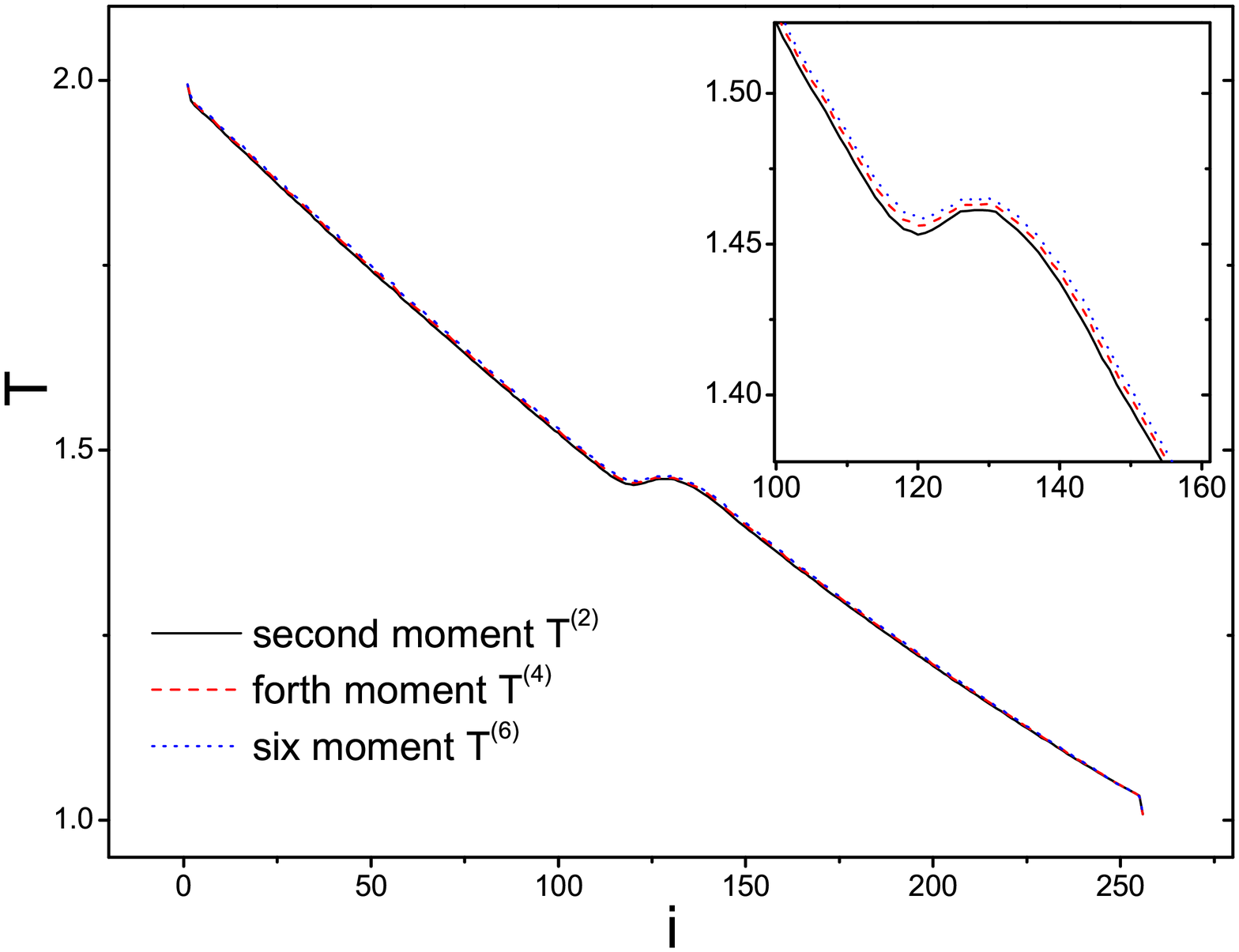}
\caption{The temperature profile of the $\phi^{4}$ lattice with an interfacial junction, whose spring constants are smoothly varied.
The distribution of the spring constants
 in the interfacial junction is the same as that for Fig.~\ref{Fig5}. Here $\lambda_{L}=\lambda_{R}=1,T_{L}=2,T_{R}=1$, and $N=256$.} \label{Fig6}
\end{figure}
So far our discussions is based on the model consisting of two segments with a single harmonic coupling, which inevitably leads to the argument that the origin of negative temperature jump comes from  the ill-defined interface of the two-segment model with a sharp discontinuity of the interfacial coupling. In what follows we propose an extended model to show that it is not the case. Actually, in a
practical consideration, the interface may be a junction which is small compared with the two sub-lattices. So we divide our system into
three regions, say, two sub-lattices and a junction. The particle number of the junction is small compared with the two sub-lattices. The spring constant of the intermediate segment varies smoothly, which is done by setting the spring constants of the intermediate junction by $k_{i}=\exp{(-(i-N/2)^{2}/50)}+1$, where $i$ represents the index of particles. The NEMD simulation results of harmonic and $\phi^{4}$ lattices are presented in Fig.~\ref{Fig5} and Fig.~\ref{Fig6}, respectively.  As we can see, negative temperature gradient still exists for both harmonic and $\phi^{4}$ lattices within the interfacial segment.

As mentioned above, a meaningful local temperature can be defined only in systems exhibiting local thermal
equilibrium. And we known that, if the system can exhibit local thermal equilibrium, the local distribution should be gaussian and all even moments can be obtained based on the second moment.  We can then use $T_{i}^{(2)}=m\langle \dot{x}_{i}^{2}\rangle$,
$T_{i}^{(4)}=m\left(\langle\dot{x_{i}}^{4}\rangle/3\right)^{1/2}$, and $T_{i}^{(6)}=m\left(\langle\dot{x_{i}}^{6}\rangle/15\right)^{1/3}$ to
define local temperature equivalently. So we carry out NEMD simulation for both the harmonic and $\phi^{4}$ lattice, and plot the
local temperature defined by the first three even moments of velocity, namely, the second, fourth and sixth moment in Fig. \ref{Fig5}
and Fig. \ref{Fig6}. To our surprise, the local temperature defined by $T^{(2)}$, $T^{(4)}$ and $T^{(6)}$ at the boundary particles agree well with each other. The deviation at the interface is not significant in comparison with that inside segments. The result indicates the local distribution is or at least very close to the Gaussian, which cannot be well distinguished by numerics and should recourse to more careful theoretical studies of local distribution in the future.
\section{SUMMARY}\label{sec4}
We have studied interfacial thermal conductance in one-dimensional inhomogeneous systems by using both theoretical calculations and MD simulations. In the weak coupling limit, theoretical calculations show that the heat current is
proportional to the square of the coupling strength in the absence of anharmonicity. A negative temperature jump between the interfacial particles occurs in both the harmonic and $\phi^{4}$ lattices. As to understand the counterintuitive observation, we have investigated the contribution of normal modes to the local temperature at the interface. It is shown that the high frequency modes make dominant contribution when the coupling strength is small, however,  the contribution of each mode is almost equivalent when the coupling strength is strong. We have confirmed the occurring of the negative temperature jump is not trivially artificial due to the integrability of the system or the sharp discontinuity of the interfacial coupling by extending the system to a model consisting of two sub-lattices and an intermediate junction for both the harmonic and $\phi^{4}$ lattices.

One should reexamine the notion of temperature as to understand the anomalous negative temperature jump, which seemingly indicats that heat flows against a local temperature gradient in a small scale. On the one hand, from the viewpoint of traditional thermodynamics, local temperature should be defined in a \textquotedblleft cell" which should be macroscopically infinitesimal but contain enough microscopic degrees of freedom. Such kind of cell is, strictly speaking, not well defined for our microscopic model due to the large atomic-scale fluctuations and the word "local" defined for a single oscillator loses its inherent meaning. On the other hand, we stress that the traditional definition of local temperature with respect to the kinetic energy of an oscillator is still in the framework of equilibrium thermodynamics. The anomalous phenomenon may partly comes from the definition as used here, which lacks a complete description of the nonequilibrium steady state. A new definition of \textquotedblleft nonequilibrium temperature" might be taken into consideration on this count~\cite{Tem1,Tem2}, especially when one take notice of the temperature profile for the middle region of the intermediate junction, which is anomalously smaller than $T_R$ as shown in Fig. \ref{Fig5}.  However, whether the concept of (local) temperature can be extrapolated beyond local equilibrium or should be modified in the nonequilibrium systems is still an open question.

\begin{acknowledgments}
The authors thank Y. Zhang, J. Wang, and H. Zhao for helpful discussioins and Xiamen Supercomputer center for using its computing facilities. This work was financially supported by NSFC No. 11047185, No. 11105112 and No. 11335006.
\end{acknowledgments}

%

\end{document}